\documentclass[journal]{IEEEtran}
\usepackage{cite}

\ifCLASSINFOpdf
  \usepackage[pdftex]{graphicx}
  \DeclareGraphicsExtensions{.pdf,.jpeg,.png}
\else
  \usepackage[dvips]{graphicx}
\fi
\usepackage{amsmath}
\usepackage{amsfonts}

\ifCLASSOPTIONcompsoc
  \usepackage[caption=false,font=normalsize,labelfont=sf,textfont=sf]{subfig}
\else
  \usepackage[caption=false,font=footnotesize]{subfig}
\fi
\usepackage{fixltx2e}

\newcommand{\z}{\mathbf{z}}
\newcommand{\x}{\mathbf{x}}
\newcommand{\q}{\mathbf{q}}
\renewcommand{\t}{\mathbf{t}}
\newcommand{\boldmu}{\mbox{\boldmath{$\mu$}}}

\newcommand{\amt}{a}

\newcommand{\A}{\mathcal{A}}
\newcommand{\D}{\mathcal{D}}

\newcommand{\ie}{\textit{i.e.}}
\newcommand{\eg}{\textit{e.g.}}
\newcommand{\eq}[1]{Eq.~(\ref{eq:#1})}
\newcommand{\fig}[1]{Fig.~\ref{fig:#1}}
\newcommand{\ig}{\includegraphics}

\usepackage{xcolor}
\newcommand{\remark}[1]{\textcolor{blue}{[\textit{#1}]}}
\renewcommand{\remark}[1]{}

\begin{document}
%
\title{Generalized likelihood ratio test detector\\
  for a
  modified replacement model target\\
  in a multivariate $t$-distributed background}
%
%
%

\author{James~Theiler
\thanks{The author is with the Space Data Science and Systems Group at Los Alamos
  National Laboratory, Los Alamos, NM 87545, USA; email: jt@lanl.gov.}
}

%
%

\markboth{}{} 
%



\maketitle

\begin{abstract}

  A closed-form expression is derived for the generalized likelihood
  ratio test (GLRT) detector of a subpixel target in a multispectral image
  whose area and brightness
  are both unknown. This expression extends a previous result (which
  assumed a Gaussian background distribution) to a
  fatter tailed elliptically-contoured (EC) multivariate $t$-distributed background.
  Numerical experiments with simulated data
  indicate that the EC-based detector outperforms the simpler Gaussian-based
  detectors, and that the relative performance of the new detector, compared to
  other EC-based detectors depends on the regime of target strength and background
  occlusion.  
\end{abstract}


%
\IEEEpeerreviewmaketitle

\section{Introduction}
\label{sect:intro}
%
%
%
%

\IEEEPARstart{T}{o obtain} a signal processing solution to the problem of
subpixel target detection in hyperspectral imagery, one requires a model for the
target signature, a model for the background distribution, and a model for how
the target signal interacts with the background.

Usually the target signature is treated as ``known.'' Although this signature
can exhibit considerable variability due to both extrinsic environmental
factors and intrinsic material factors\cite{Theiler19}, the extent of this
variability is in most cases assumed to be known, and is very often treated
as a constant.

The background is not ``known'' in the same \textit{a priori} sense
that the target is known, but it is modeled by a probability
distribution that can often be estimated directly from the
hyperspectral imagery in which the target is sought.  The Gaussian is
perhaps the most popular choice; it is parameterized by a mean vector
$\boldmu$ and covariance matrix $R$.  The multivariate $t$
distribution is an attractive choice: like the Gaussian, it requires a
vector-valued mean and a matrix-valued covariance, but it takes one
further parameter, the scalar $\nu$, which characterizes the tail of
the distribution.  The multivariate $t$ is an elliptically contoured
(EC) distribution, because the contours of constant probability
density are ellipsoids centered at $\boldmu$ (indeed, the same
ellipsoids that characterize contours of the Gaussian distribution),
but it exhibits fatter tails than a Gaussian.  Manolakis
\textit{et~al.}\cite{Manolakis01,Marden04} have argued that these
fatter-tailed EC distributions are appropriate for modeling
hyperspectral background variability.

Several models have been described that characterize how the target
interacts with the background.  For an opaque target that occupies a full pixel,
there is no interaction, and the detector is straightforward to the point of trivial.
Here, we have a simple hypothesis testing problem:
\begin{align}
  H_o&: \x = \z \\
  H_1&: \x = \t
\end{align}
where $\x$ is the observed spectrum at a pixel, $\z$ is the background at that pixel (\ie, what the spectrum would be if no target were present), and $\t$ is the target spectrum.  There are no nuisance parameters in this model, so the optimal detector is given by the likelihood ratio $\D(\x) = p_t(\x)/p_z(\x)$, where $p_z$ and $p_t$ are the probability density functions associated with the background and target, respectively.  We can interpret $p_z(\x)$ is the likelihood of observing $\x$ in a background pixel, and $p_t(\x)$ as the likelihood of observing $\x$ in a target pixel.

For subpixel targets (or for models that incorporate adjacency effects, or include translucency, such as gas-phase plume targets\cite{Theiler20igarss}), there is an interaction of target with background.  The most common models are the additive and replacement models, though Vincent and Besson\cite{Vincent20c} have recently suggested a hybrid ``modified replacement'' model that generalizes both of them, though at the cost of introducing a second parameter:

\begin{tabular}{rll}
  &&\\
  $H_1$:& $\x = \z + \alpha\t$ & \mbox{\it Additive model}\\
  $H_1$:& $\x = (1-\alpha)\z + \alpha\t$ & \mbox{\it Replacement model}\\
  $H_1$:& $\x = \beta\z + \alpha\t$ & \mbox{\it Modified replacement model}\\
  &&
\end{tabular}

The additive model corresponds to a target whose strength depends
on~$\alpha$, where ``strength'' in this case might correspond to
intrinsic brightness (or reflectance) of the target, depending on
linear factors that might include illumination, temperature (for
infrared imaging), concentration (especially for gas-phase plumes),
etc.  The replacement model treats the subpixel target as opaque, and
having an area (relative to the pixel size) of~$\alpha$; thus its
contribution to the observed signal is proportional to~$\alpha$ but
the contribution of the background is correspondingly diminished by a
factor of $1-\alpha$.  The modified replacement model is attractive in
that it can account both for the diminishing background contribution
($\beta<1$) due to an opaque (or partially opaque) target, and for
target strength variability due to more intrinsic brightness effects.
Solid materials often exhibit spectral variability due, for instance,
to powder grain size, and although that variability may be somewhat
complex, it is often seen in practice that the dominant effect is in
the overall magnitude of the reflectance of the material.

Unlike the full-pixel case, the alternative hypotheses (the $H_1$'s)
for these target-background interaction models involve unknown, or
nuisance, parameters: $\alpha$, or $\alpha$ and $\beta$. For this
reason, they are not simple but \emph{composite} hypothesis testing
problems.  There are a number of approaches for dealing with composite
hypothesis testing problems: Bayes factor\cite{Lehmann05}, penalized
likelihood\cite{Chen98,Vexler10}, and clairvoyant
fusion\cite{Schaum10,Schaum16b} among them.  The most popular, and
usually quite effective (albeit not always
optimal\cite{Theiler12spie}), is the generalized likelihood ratio test
(GLRT), which is based on maximizing the likelihoods used in the
likelihood ratio test, or equivalently on employing maximum likelihood
estimates of the nuisance parameters in the likelihood ratio test.  

Using the GLRT, closed-form expressions have been identified for
a variety of target models on a variety of background distributions.

For an additive target on a Gaussian background, the adaptive matched
filter (AMF) is the appropriate detector\cite{Reed74,Robey92}; indeed, it is
provably optimal as the uniformly most powerful (UMP)
detector\cite{Lehmann05}.  For solid subpixel targets, a replacement
model is appropriate. The target signal is proportional to the
fraction $\amt$ of the pixel that the target covers, but the
background is occluded by that same fraction.  For Gaussian
background, this leads to the finite target matched filter
(FTMF)\cite{Schaum97}.  Finally, a generalization of additive and
replacement is given by the modified replacement model: here both the
target and the background scale, and the Gaussian GLRT is given by
Vincent and Bresson\cite{Vincent20c}. In~[\citenum{Vincent20c}],
both one-step and two-step variants are derived, and the one-step
variant is called SPADE (Sub-Pixel Adaptive DEtection); the interest
in this paper is with the two-step variant, which I will call 2SPADE.
\remark{The authors\cite{Vincent20c} call this two-step variant the ``modified FTMF,''
  and since they first derived it, they arguably have naming rights,
  but I think it is clearer to just call a spade a spade.}

These are two-step detectors; the mean and covariance are computed
from a (large) sample of off-target pixels, and are treated as exact
and fixed for the pixels under test.  For global methods (in which a
single mean and covariance is estimated for the whole image) and even
for semi-local methods (in which the mean is computed locally, but a
single global covariance is estimated), the sample size will be large
and the two-step methods are appropriate.   In practice, we may not know if a set of pixels are truly target-free, and in that case some level of contamination may occur.  As long as the contaminating targets are rare and/or weak, the effect of this contamination will usually be small\cite{Theiler06}.

But for local methods, in which both the mean and covariance are recomputed
in a (small) moving window, the one-step methods are potentially better
detectors because they account for the statistical imprecision in the estimates
of mean and covariance.  The one-step derivations are more difficult to derive, but a
one-step AMF is given by
Kelly\cite{Kelly89}, and one-step replacement (ACUTE) by Vincent and
Bresson\cite{Vincent20a}, and one-step modified replacement
(SPADE)\cite{Vincent20c}.  In all three of these cases, as the sample size
becomes large, the one-step detector approaches the two-step detector.

The extension of these two-step detectors to elliptically-contoured multivariate
$t$-distributed backgrounds has been developed for the
additive\cite{Theiler08igarss} and for the
replacement\cite{Theiler18igarss} models.  In this exposition, the
two-step modified replacement model will be extended to an
elliptically contoured background, here called EC-2SPADE.

Also, beyond Gaussian and multivariate-$t$ are non-parametric methods,
such as NP-AMF\cite{Matteoli20glrt}, which enable far more flexible
modeling of the background.

\begin{table}
  \caption{Summary of detectors for various target models on various background
  distributions. The EC-2SPADE detector is derived in this paper.}
  \label{tab:summary}
  \begin{tabular}{llll}
    \multicolumn{2}{c}{Target Model} & \multicolumn{2}{c}{Background distribution} \\
    Model Name & Expression & Gaussian & multivariate $t$ \\
    \hline\\
    Additive & $\x = \z + \alpha\t$ & 
    AMF\cite{Reed74,Robey92} & EC-AMF\cite{Theiler08igarss} \\
    Replacement & $\x = (1-\alpha)\z + \alpha\t$ & 
    FTMF\cite{Schaum97} & EC-FTMF\cite{Theiler18igarss} \\
    Modified & $\x = \beta\z + \alpha\t$ & 
    2SPADE\cite{Vincent20c} & EC-2SPADE\\
    \hline
\end{tabular}
\end{table}

\section{Set up the problem}
\subsection{Background model}
In the absence of target, the background is assumed to be distributed as a
multivariate $t$ distribution.  A background pixel spectrum is denoted
$\z\in\mathbb{R}^d$ where $d$ is the number of spectral channels in the hyperspectral
imagery.  Here,
\begin{equation}
  p_z(\z) = c \left[ (\nu-2) + \A(\z)\right]^{-(d+\nu)/2}
\end{equation}
where $c$ is a normalizing constant, 
$\nu$ is a scalar
parameter that characterizes the tail of the distribution, and
\begin{equation}
  \A(\z) = (\z-\boldmu)'R^{-1}(\z-\boldmu)
\end{equation}
is a squared Mahalanobis distance.  Here, $\boldmu$ and $R$ are the mean and covariance of the background, and the fact that $p(\z)$ depends on $\z$ through $\A(\z)$ ensures that the distribution is elliptically contoured.  Note that in the limit as $\nu\to\infty$, the distribution becomes Gaussian.  For $\nu \le 2$, the distribution is so fat-tailed
that the second moment does not exist.

\subsection{Target-background interaction model}
Under the null hypothesis (which is that the target is not present in the given pixel),
the observed spectrum is the background.  The alternative hypothesis is that the
target is present.  The target has a known
signature $\t$, but what is observed is a linear combination of target and background.
Thus:
\begin{align}
  H_o&: \x = \z \\
  H_1&: \x = \beta\z + \alpha\t
\end{align}
with constraints $0\le\beta\le 1$ and $0\le\alpha$.

\section{GLRT solution}
Because this is a two-step solution, we will begin with the assumption that we have an adequate estimate of $\boldmu$, $R$, and $\nu$, usually obtained from a large number of background pixels. 

Under the hypothesis that a target is present, we have $\x = \beta\z + \alpha\t$, where $\beta$ and $\alpha$ are unknown.  Then
\begin{equation}
  \z = \frac{\x - \alpha\t}{\beta}
\end{equation}
and the probability distribution associated with observation $\x$ is given by
\begin{align}
  p_x(\alpha,\beta;\,\x) &= p_z(\z)\left|\frac{d\z}{d\x}\right|
  = p_z\left(\frac{\x - \alpha\t}{\beta}\right)\cdot\beta^{-d}\\
  &= c \beta^{-d} \left[ (\nu-2) + \A\left(\frac{\x - \alpha\t}{\beta}\right)\right]^{-(d+\nu)/2} \label{eq:px}
\end{align}
The GLRT detector the is based on the likelihood ratio, maximized over the
nuisance parameters; specifically
\begin{equation}
  \D(\x) = \frac{\max_{\alpha,\beta}p_x(\alpha,\beta;\,\x)}{p_x(0,1; \x)}
  = \frac{p_x(\widehat\alpha,\widehat\beta;\,\x)}{p_z(\x)}
  \label{eq:detector}
\end{equation}
where $\widehat\alpha$ and $\widehat\beta$ are the values (they are both functions of $\x$) that maximize $p_x(\alpha,\beta;\,\x)$.

We will begin with maximization over~$\alpha$.  This occurs at
\begin{align}
  \widehat\alpha &= \mbox{argmax}_\alpha~ p_x(\alpha,\beta;\,\x) \\
  &= \mbox{argmax}_\alpha~ \A\left(\frac{\x - \alpha\t}{\beta}\right) \\
  &= \mbox{argmax}_\alpha~ (\x - \alpha\t - \beta\boldmu)'R^{-1}(\x - \alpha\t - \beta\boldmu) \\
  &= \frac{\t'R^{-1}(\x-\beta\boldmu)}{\t'R^{-1}\t}  \label{eq:hat-alpha}
\end{align}
Observe that
\begin{align}
  \frac{\x-\widehat\alpha\t}{\beta} - \boldmu = \frac{1}{\beta}\left(I - \frac{\t\t'R^{-1}}{\t'R^{-1}\t}\right)(\x-\beta\boldmu)
\end{align}
Thus, if we write
\begin{equation}
  Q = \left(I - \frac{\t\t'R^{-1}}{\t'R^{-1}\t}\right)'R^{-1}
  \left(I - \frac{\t\t'R^{-1}}{\t'R^{-1}\t}\right)
  \label{eq:Q}
\end{equation}
then we have
\begin{align}
  \A\left(\frac{\x - \widehat\alpha\t}{\beta}\right) &=
  \frac{(\x-\beta\boldmu)' Q (\x-\beta\boldmu)}{\beta^2} \\
  &= a + b\beta^{-1} + c\beta^{-2} = q(\beta)
\end{align}
where $q(\beta)$ is a scalar quadratic expression in $\beta^{-1}$ with
\begin{align}
  a &= \boldmu' Q \boldmu, \\
  b &= -2\boldmu' Q \x, \\
  c &= \x' Q \x.
\end{align}

With $\alpha=\widehat\alpha$, our expression for likelihood in \eq{px} becomes
\begin{align}
  p_x(\beta;\,\x) &= p_x(\widehat\alpha(\beta,\x),\beta;\,\x) \\
  &= c \beta^{-d} \left[ (\nu-2) + q(\beta) \right]^{-(d+\nu)/2}
\end{align}
Writing the log likelihood (minus a constant) for hypothesis $H_1$, we have
\begin{align}
  L_1(\beta;\,\x) &= \log p_x(\beta;\,\x) - \log c \\
  &= -d\log\beta -\frac{d+\nu}{2}\log\left[(\nu-2) + q(\beta)\right]
  \label{eq:loglike}
\end{align}
To maximize this log likelihood we take the derivative with respect to~$\beta$
and set the result to zero:
\begin{equation}
  0 = \frac{\partial}{\partial\beta} L_1(\beta;\,\x)
  = -d\beta^{-1} -\frac{d+\nu}{2}\frac{\frac{\partial}{\partial\beta} q(\beta)}{(\nu-2) + q(\beta)}
\end{equation}
Multiplying both sides
by $(\nu-2)+q(\beta)$, we obtain
\begin{align}
  0 &= -d\beta^{-1}[(\nu-2)+q(\beta)] -\frac{d+\nu}{2}[-b\beta^{-2} -2c\beta^{-3}]
\end{align}
Now, multiply both sides by $-\beta^3/\nu$:
\begin{align}
  0 &= \frac{d(\nu-2+a)}{\nu}\beta^2 +
       \frac{db}{\nu}\beta + dc + \frac{d+\nu}{2\nu}[-b\beta -2c] \\
  &= A\beta^2 + B\beta + C
\end{align}
which is a quadratic equation in $\beta$ and can be solved in closed-form.
Here,
\begin{align}
  A &= d + \frac{d(a-2)}{\nu} \\
  B &= -\frac{b}{2} + \frac{db}{2\nu} \\
  C &= -c 
\end{align}
from which
\begin{equation}
  \widehat\beta = \mbox{min}\left(\frac{-B+\sqrt{B^2-4AC}}{2A}, 1\right)
  \label{eq:hat-beta}
\end{equation}
Note that since $A>0$ and $C<0$, we can be sure that $\widehat\beta>0$.
We can further observe that $\widehat\beta=1$ whenever $-C\ge(A+B)$.
  \remark{For the Gaussian case (below, $\nu\to\infty$) with $\boldmu=0$ (a very special
  case), this corresponds to $\x'R^{-1}\x \ge d + (\q'\x)^2$.}

Finally, given $\widehat\alpha$ from \eq{hat-alpha} and $\widehat\beta$ from
\eq{hat-beta}, the GLRT target detector for the modified replacement model
is given by \eq{detector}.

\subsection{Special case: $\nu\to\infty$}
In the $\nu\to\infty$ limit, the EC distribution becomes Gaussian.  The expressions for $A$, $B$, and $C$ become
\begin{align}
  A &= d \\
  B &= -b/2 = \boldmu'Q\x \\
  C &= -c = -\x'Q\x
\end{align}
which is consistent with Eq(6) from Ref.~[\citenum{Vincent20c}].

\subsection{Remark on $Q$}
Note that \eq{Q} can equivalently be written
\begin{equation}
  Q = R^{-1/2}\left(I - \frac{R^{-1/2}\t\t'R^{-1/2}}{\t'R^{-1}\t}\right)R^{-1/2}
\end{equation}
which can be interpreted as a projection operator sandwiched between two whitening operators.  If we consider the matched filter vector
\begin{equation}
  \q = R^{-1}\t/\sqrt{t'R^{-1}\t}
\end{equation}
then we have another expression for $Q$ given by
\begin{equation}
  Q = R^{-1} - \q\q'
\end{equation}

\section{Clairvoyant detector}

If $\alpha$ and $\beta$ were known, then the so-called clairvoyant
detector\cite{Kay98} provides optimal detection, but it is an odd
scenario to know the strength of the target without knowing whether or
not the target is present.  Still, the clairvoyant provides a useful
upper bound on the performance of a target detector, and shows the
penalty paid by replacing the true (but unkonwn) $\alpha$ and $\beta$
with the estimates $\widehat\alpha$ and $\widehat\beta$ which are
recomputed for each pixel.  The detector is given by any monotonic
transform ($h$) of the likelihood ratio:
\begin{equation}
  \D(\alpha,\beta,\x) = h\left( \frac{p_x(\alpha,\beta;\,\x)}{p_z(\x)} \right).
\end{equation}
If we let
\begin{equation}
  h(w) = \frac{\exp\left(\frac{2}{d+\nu}\left[\log w + d\log\beta\right]\right)-1}{\nu-2},
\end{equation}
\remark{Note that $h(w)$ depends on $\beta$; this is fine for deriving clairvoyant
  detectors, because $\beta$ is a constant, independent of $\x$.  But we cannot,
  for example, use \eq{clairvoyant} to derive $\widehat\beta$, and we will not
  obtain a GLRT detector by plugging the $\widehat\alpha,\widehat\beta$ into \eq{clairvoyant}.}
then
\begin{equation}
  \D(\alpha,\beta,\x) = \frac{1}{1+\displaystyle\frac{\A(\x)}{\nu-2}}
  \left[\A\left(\frac{\x-\alpha\t}{\beta}\right)-\A(\x)\right]
    \label{eq:clairvoyant}
\end{equation}
is a clairvoyant detector.  In the $\nu\to\infty$ limit, this is simply
\begin{equation}
  \D(\alpha,\beta,\x) = \A\left(\frac{\x-\alpha\t}{\beta}\right)-\A(\x).
\end{equation}

\section{Numerical illustration}

\fig{roc} shows ROC curves, illustrating detector performance for
$\alpha=0.2$ and a range of values of $\beta$.  One very general
trend is that detection becomes harder as $\beta$ gets larger, and
more background signal is mixed in with the target signal.

The performance is computed using simulated data, and since that
simulated data is based on an elliptically-contoured (EC) multivariate
$t$ background, it is not surprising to observe that the EC-based
algorithms (shown with solid lines) generally outperform their Gaussian
counterparts (dashed lines).

We see for small values of $\beta$ that the 2SPADE and
EC-2SPADE are the best detectors, with EC-2SPADE outperforming
2SPADE by a considerable margin.  As $\beta$ increases toward
$1-\alpha$, the FTMF and EC-FTMF algorithms begin provide the best
performance, with EC-FTMF outperforming FTMF. This cross-over in ROC
curve performance between 2SPADE and FTMF was also observed by Vincent
and Besson\cite{Vincent20c}.  Finally, as $\alpha\to 1$, the EC-AMF
exhibits the best performance.

Similar curves are seen in \fig{roc-a6}, but here $\alpha=0.6$ which
provides more examples with $\alpha+\beta>1$.  For $0.3 \le \beta \le 0.5$, we see
that EC-FTMF outperforms EC-2SPADE. For $\beta=0.6$, EC-2SPADE is best, but for $\beta\ge 0.7$, EC-AMF outperforms EC-2SPADE.
\remark{Not shown here, but for $\beta\le 0.2$, EC-2SPADE again wins; for such small
  $\beta$, near perfect detection is possible and is achieved by EC-2SPADE.}
On the other hand, in none of the cases here is EC-2SPADE the worst detector, and it
is never as bad as EC-AMF at small $\beta$ or EC-FTMF at large~$\beta$.

Because we are comparing performance to the optimal clairvoyant
detector, we can see that, at their points of optimailtiy, the EC-FTMF
(at $\beta=1-\alpha=0.8$) and EC-AMF ($\beta=1.0$) detectors are very
nearly optimal in their performance.  By contrast the EC-2SPADE
detector does not seem to approach the performance of the clairvoyant
detector.  Fitting two parameters instead of one seems to
incur a performance penalty.

\newcommand{\lig}[2]{\parbox{0.25\textwidth}{\centerline{#1} \ig[width=0.25\textwidth]{#2}}}
\begin{figure*}
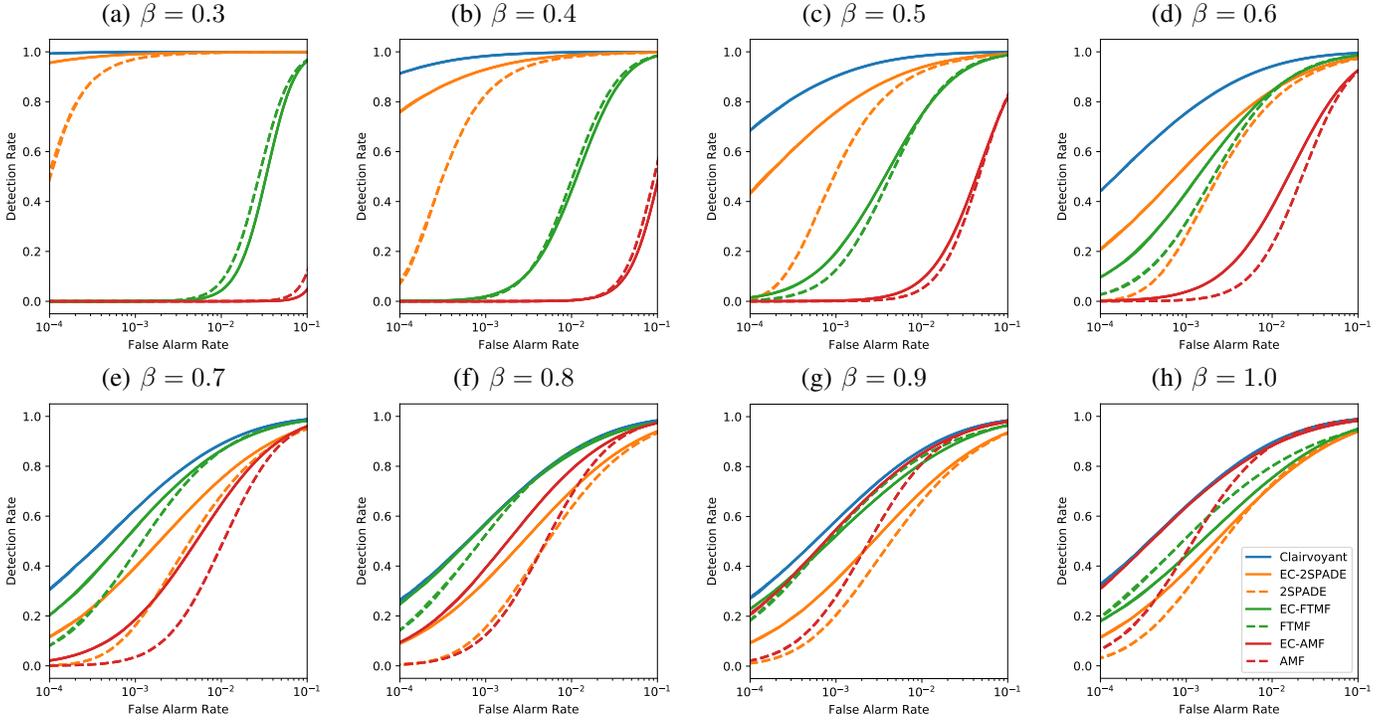

  \centerline{
    \lig{(a) $\beta=0.3$}{ROC-d10-nu10-mu2-t15-a02-b03x}
    \lig{(b) $\beta=0.4$}{ROC-d10-nu10-mu2-t15-a02-b04x}
    \lig{(c) $\beta=0.5$}{ROC-d10-nu10-mu2-t15-a02-b05x}
    \lig{(d) $\beta=0.6$}{ROC-d10-nu10-mu2-t15-a02-b06x}
  }
  \centerline{
    \lig{(e) $\beta=0.7$}{ROC-d10-nu10-mu2-t15-a02-b07x}
    \lig{(f) $\beta=0.8$}{ROC-d10-nu10-mu2-t15-a02-b08x}
    \lig{(g) $\beta=0.9$}{ROC-d10-nu10-mu2-t15-a02-b09x}
    \lig{(h) $\beta=1.0$}{ROC-d10-nu10-mu2-t15-a02-b10x}
  }
  \caption{ROC curves showing detector performance for a range of
    $\beta$ values from 0.3 to 1.0.  For these simulations, $N=10^7$
    matched-pair samples are generated; for each pair, one is without
    target ($\x=\z$) and one includes target ($\x=\beta\z+\alpha\t$).
    The data are simulated from a multivariate $t$ distribution, with
    $d=10$ and $\nu=10$, mean $\mu = [2,2,\ldots,2]'$, and unit
    covariance $R=I$.  The target signature is $\t = \mu +
    [T,0,\ldots,0]$, with $T=15$, and the target strength for all
    these simulations was $\alpha=0.2$.  (Very roughly speaking, the
    target is ``three sigmas'' away from the background mean.) Note
    that three curves are drawn for each case in order to provide a
    sense of trial-to-trial variability; for the curves here, however,
    that variability rarely exceeds the linewidth on the plots.}
  \label{fig:roc}
\end{figure*}
\begin{figure*}
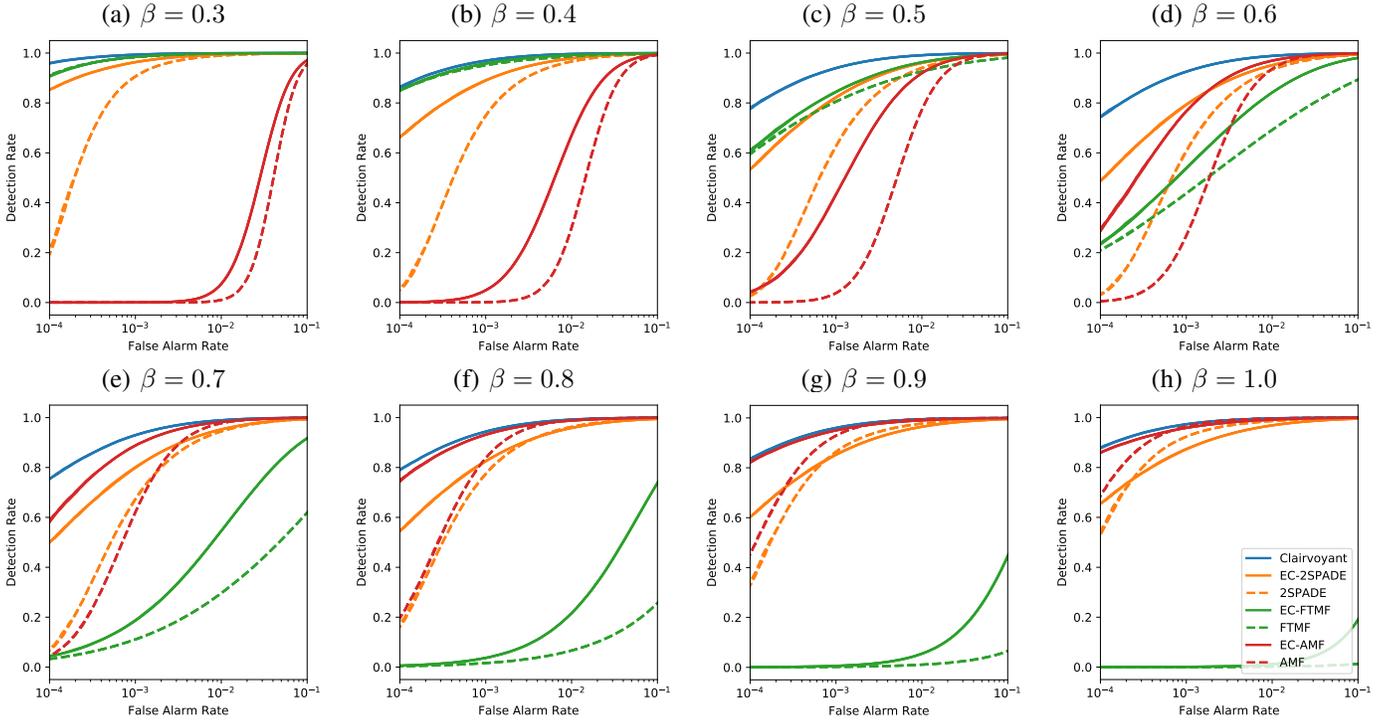

  \centerline{
    \lig{(a) $\beta=0.3$}{ROC-d10-nu10-mu2-t5-a06-b03x}
    \lig{(b) $\beta=0.4$}{ROC-d10-nu10-mu2-t5-a06-b04x}
    \lig{(c) $\beta=0.5$}{ROC-d10-nu10-mu2-t5-a06-b05x}
    \lig{(d) $\beta=0.6$}{ROC-d10-nu10-mu2-t5-a06-b06x}
  }
  \centerline{
    \lig{(e) $\beta=0.7$}{ROC-d10-nu10-mu2-t5-a06-b07x}
    \lig{(f) $\beta=0.8$}{ROC-d10-nu10-mu2-t5-a06-b08x}
    \lig{(g) $\beta=0.9$}{ROC-d10-nu10-mu2-t5-a06-b09x}
    \lig{(h) $\beta=1.0$}{ROC-d10-nu10-mu2-t5-a06-b10x}
  }
  \caption{ROC curves showing detector performance for a range of
    $\beta$ values from 0.3 to 1.0.  Same as \fig{roc}, but here
    $\alpha=0.6$ (three times larger) and $T=5$ (three times weaker),
    so there is still a roughly three-sigma distance from the target
    to the mean.}
  \label{fig:roc-a6}
\end{figure*}

\section{Discussion}

Applied to simulated data, tailored to the assumptions of the new detector,
the new detector performs well.  Further work will be needed to examine how
well this detector behaves in the ``wild,'' with real targets in real imagery.

Introducing, and optimizing over, more parameters does lead to a more flexible
detector, but not necessarily a more powerful one.  We saw that imposing an assumption
(\eg, that $\beta=1-\alpha$) that reduced the number of free parameters led to
improved detection performance even when the assumption was only approximately
true.



%

\bibliography{ecmrm}
\bibliographystyle{IEEEtran}

%
%

%







\end{document}